\documentclass[conference]{IEEEtran}
\usepackage{caption}
\IEEEoverridecommandlockouts
\usepackage{cite}
\usepackage{amsmath,amssymb,amsfonts}
\usepackage{algorithmic}
\usepackage{graphicx}
\usepackage{textcomp}
\usepackage{xcolor}
\usepackage{float}
\usepackage{bm}
\usepackage{subcaption}
\usepackage{stfloats}
\usepackage{siunitx}
\usepackage{array} 
\usepackage{soul}
\usepackage{tikz}
\usetikzlibrary{arrows.meta, shapes, positioning}

\def\BibTeX{{\rm B\kern-.05em{\sc i\kern-.025em b}\kern-.08em
    T\kern-.1667em\lower.7ex\hbox{E}\kern-.125emX}}
\begin{document}

\title{Obstacle Detection at Level Crossings under Adverse Weather Conditions --- a Survey}

\author{Chenyang~Yan,~\IEEEmembership{Graduate Student Member,~IEEE,}
        Mats~Bengtsson,~\IEEEmembership{Senior Member,~IEEE}%
\thanks{The authors are with the School of Electrical Engineering and Computer
Science, KTH Royal Institute of Technology, Stockholm, Sweden (e-mail: \{chya, matben\}@kth.se).}%
\thanks{Funded by the European Union through Trafikverket. Views and opinion expressed are however those of the author(s) only and do not necessarily reflect those of the European Union or the Europe’s Rail Joint Undertaking. Neither the European Union nor the granting authority can be held responsible for them. The project FA6 FutuRe is supported by the Europe’s Rail Joint Undertaking and its members.}
}

\maketitle

\begin{abstract}
Level crossing accidents remain a significant safety concern in modern railway systems, particularly under adverse weather conditions that degrade sensor performance. This review surveys the state-of-the-art in sensor technologies and fusion strategies for obstacle detection at railway level crossings, with a focus on robustness, detection accuracy, and environmental resilience. Individual sensors such as inductive loops, cameras, radar, and LiDAR exhibit unique advantages but suffer from inherent trade-offs—ranging from material or visibility dependency to resolution limitations in harsh environments. We analyze each modality’s working principles, weather-induced vulnerabilities, and mitigation strategies including signal enhancement and machine learning-based denoising. Furthermore, we explore multi-sensor fusion approaches—categorized into data-, feature-, and decision-level architectures—that integrate complementary sensor information to enhance detection reliability and fault tolerance. The survey concludes by outlining future research directions, including the development of adaptive fusion algorithms, real-time processing pipelines, and weather-resilient datasets to support the deployment of intelligent, fail-safe detection systems for railway safety.
\end{abstract}


\section{Introduction}
Railways have always been a crucial mode of transportation in Europe. According to the latest edition of "Key Figures on European Transport" published by the European Union, in 2022, trips between EU countries reached 7.3 billion, covering a total distance of 372 billion kilometers. Additionally, international transportation mileage was 21 billion kilometers. On a per capita basis, domestic travel averaged 833 kilometers per inhabitant, while international travel averaged 47 kilometers per inhabitant \cite{KeyFig2023}. However, in 2021, there were still 1,389 significant accidents reported, along with an estimated economic cost of around 3.2 billion EUR per annum, marking an increase compared to 2020. The rise in “external” accidents in 2021 was mainly due to an increase in level crossing accidents. In 2021, around 96,000 level crossings were reported in the EU-27 Member States, with passive level crossings accounting for around 42\% of the total. These level crossings are usually equipped with a St Andrew’s cross traffic sign but do not provide any active warning to road users \cite{Safty2023}.

Level crossing safety is one of the most important topics in the railway industry and has been discussed as early as the 19th century. Back then, there were no techniques supporting the safety issue and most of the tasks were done by manual operation and construction of guardrails and warning signs~\cite{networkrail2024}. Since the 20th century, with technological improvements, different kinds of sensors have been used to detect obstacles at level crossings, enhancing the safety of trains and pedestrians, as well as protecting property. Obstacle detection systems are particularly relevant for low-traffic railway lines, where replacing level crossings with overpasses or tunnels, although inherently safer, may not be economically feasible due to their high construction and maintenance costs.

An ideal obstacle detector for level crossings should meet the following requirements: 1) enhance safety at level crossings; 2) cause minimal delays to train operations or road traffic; 3) be cost-effective in terms of installation, operation, and maintenance; 4) be practical to use and maintain \cite{ADLittle:2006}. It should be noted that requirement 2) also necessitates a fast reaction time from the level crossing detector, and requirement~4) implies that the detector should perform well in all weather conditions and temperatures \cite{railengineer2024}.

From an economic standpoint, the cost of the hardware is often significantly lower than that of installation and maintenance—particularly when servicing requires disruption of rail traffic. Therefore, systems that can be installed beside the rail bank, avoiding interference with active railway tracks, offer substantial advantages in terms of both cost and operational feasibility.

Obstacles of concern can be categorized based on their size and material. Common obstacles include humans, animals, metal objects (such as other vehicles using the level crossing), and other intruders like falling rocks, or goods following off a truck. In contrast, low-density, small, or soft items should be considered harmless and ignored by the detector. Examples of such items include umbrellas (when not held by a person), cardboard boxes, newspapers, fog, falling snow, or heavy rain ~\cite{railengineer2024}.

It is also relevant to note that the formal requirements for what constitutes a safety-critical obstacle may vary between countries. For example, national railway authorities may specify different minimum detection criteria and obstacle size thresholds depending on local risk assessments and operational procedures. In Sweden, for instance, draft regulations suggest that an obstacle detection system must reliably detect a human-sized object on the track, while disregarding low-risk items such as paper debris or snow accumulation. These differences in regulatory expectations highlight the need for adaptable detection strategies that can be tuned to comply with specific national standards.

Under different weather conditions and for various detection targets, sensor performance can vary significantly. Some commonly used sensors, such as inductive loops, are limited to detecting metal objects. As a result, they are widely used to detect vehicles but are ineffective for detecting pedestrians \cite{gajda2001vehicle,Urbantraffic,desai2014evaluation}.
Another example is visual-based sensors, such as cameras and thermal cameras. These sensors collect graphic information about obstacles, allowing them to distinguish the figure of a human. However, they are easily affected by weather conditions and can raise privacy concerns. Moreover, visual data does not provide material information, making it difficult to assess the risk level \cite{ristic2021review}. To overcome these issues, other sensors have been studied, including radar, LiDAR, optical beams, and also sensor combinations. 

In this review, we aim to list and compare a series of sensor based obstacle detection technologies used at level crossings. We will focus not only on the hardware setups but also on the signal processing methods employed by different sensors. Focusing on the current practical problems, we also examine the potential performance of these sensors when used in extreme weather conditions or for detecting difficult obstacles. This analysis will provide insights for the future generation of railway obstacle detection systems.

\section{Sensor Technologies for Obstacle Detection}

\subsection{Inductive Loops}
Inductive loops were developed in the early 1960s and have been used for over 60 years. They are still widely employed to detect vehicles and trains today~\cite{GAJDA201257,singhal2022sensor}. Their operation is grounded in electromagnetic principles, where detectability is predominantly governed by an object's metallic content. When a vehicle or train enters the loop field, its body and frame provide a conductive path for the magnetic field. This produces a loading effect, causing the loop inductance to decrease. The decreased inductance leads to an increase in the resonant frequency from its nominal value. If the frequency change exceeds the threshold set by the sensitivity setting, the loop processor outputs a detection signal \cite{klein2006traffic}. However, non-metallic objects, such as pedestrians, animals, or lightweight plastic debris, typically fail to induce sufficient electromagnetic disturbance, rendering them effectively invisible to inductive loop systems.

Typically, the loop field consists of one or more turns of insulated loop wire typically installed below the upper surface of a sleeper \cite{petrov2011loop}. The loop wire can be made of reinforcing steel mesh or rebar. Since the loop wires are set underground and the area of the loop field defines the detection zone, this method is minimally affected by environmental conditions such as heavy rain or fog. Additionally, inductive loops have a relatively long lifespan, with a mean time between failures of approximately eight years \cite{ecmfrance2024}. However, inductive loops are not entirely satisfactory for level crossing obstacle detection for several reasons: they can only detect objects with a sufficient amount of metal; the installation and maintenance of loop sensor arrays buried in the railroad will disrupt train operations; limited installation areas may not cover the entire level crossing; and there is a lack of redundancy and system-level performance monitoring \cite{hilleary2012radar}.

\subsection{Cameras}
\begin{figure}[h]
    \centering
    \begin{subfigure}{0.3\textwidth}
        \centering
        \includegraphics[width=\linewidth]{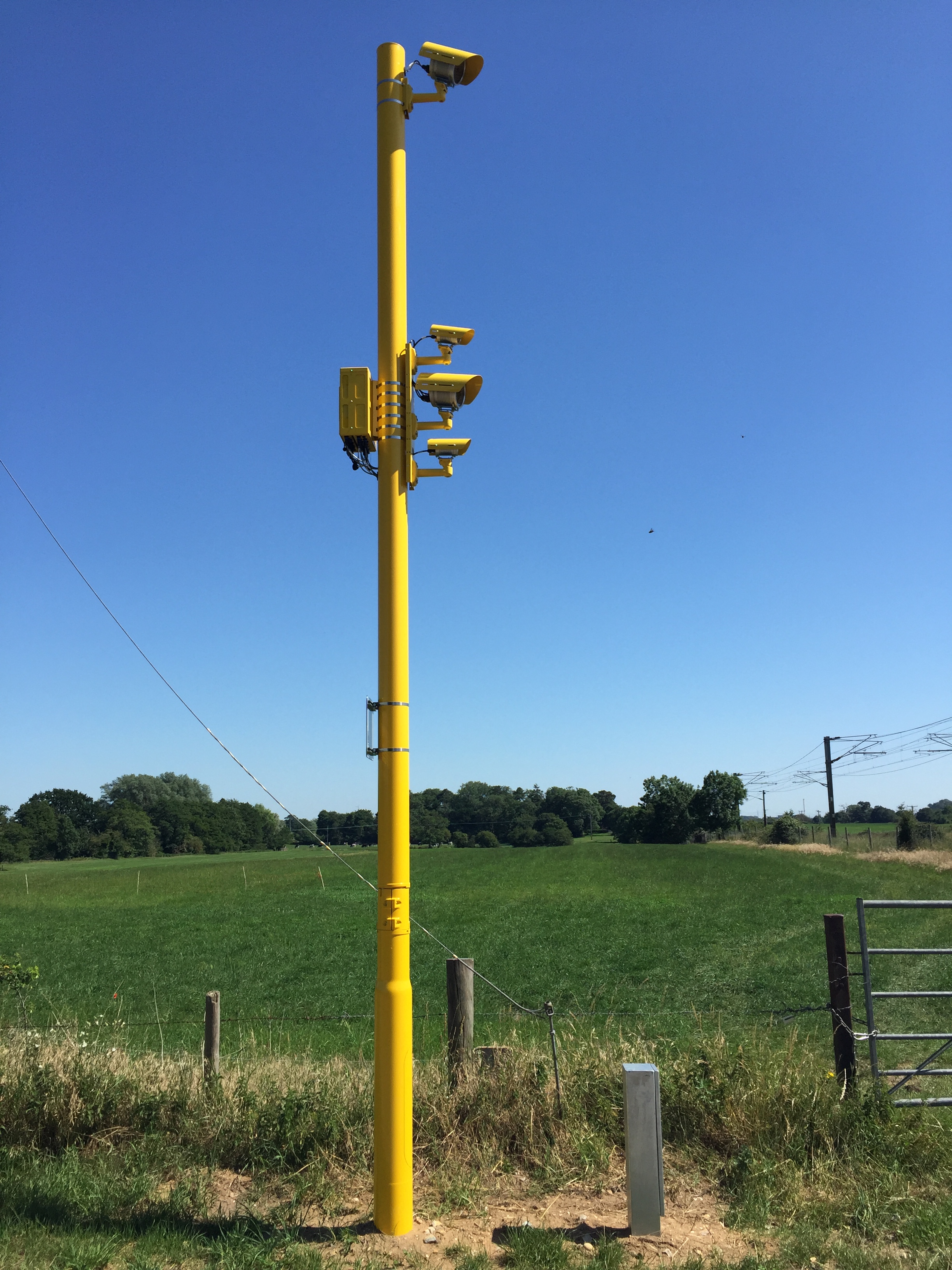}
        \caption{Surveillance camera at level crossing}
    \end{subfigure}%
    \hfill
    \begin{subfigure}{0.3\textwidth}
        \centering
        \includegraphics[width=\linewidth]{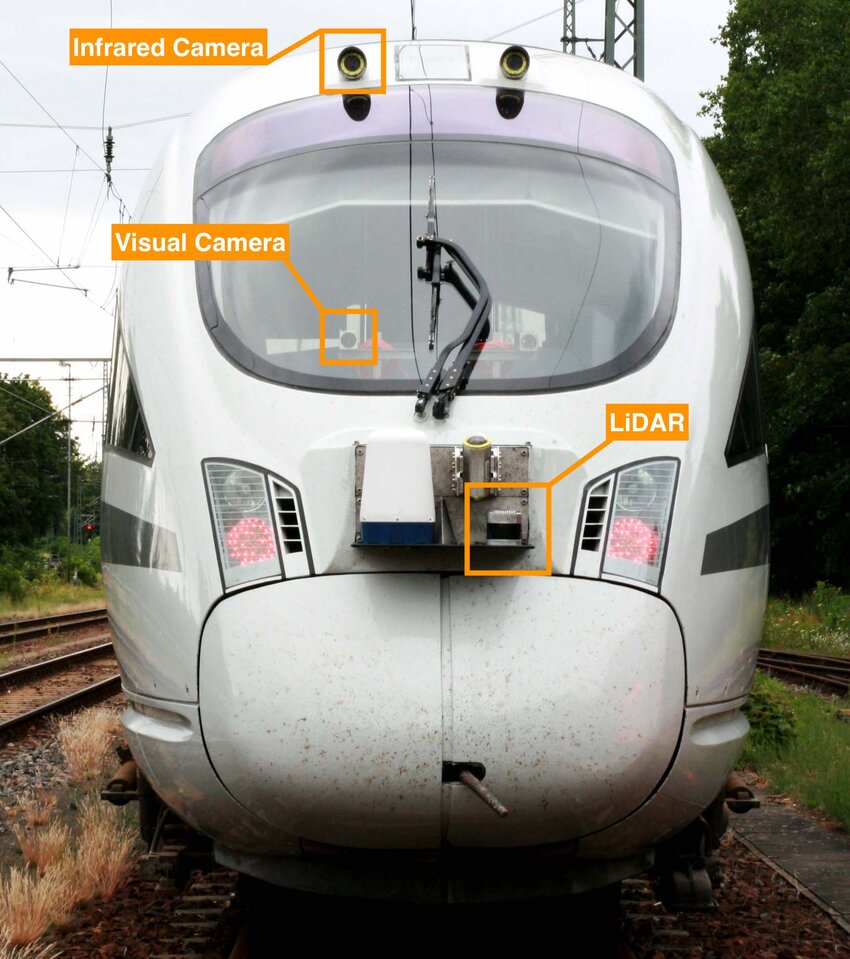}
        \caption{On-board visual sensors}
    \end{subfigure}
    \caption{Examples of camera deployment in railway systems: (a) fixed surveillance camera at a level crossing, (b) on-board sensors mounted on a train \cite{denzler2022multi}.}
    \label{cameras}
\end{figure}

In railway obstacle detection systems, several types of cameras are used to capture visual information. These include RGB cameras, stereo cameras, thermal cameras, and night vision cameras. Based on their installation location, cameras are categorized as either on-board (mounted on trains) or stationary (installed at level crossings). These visual sensors support both traditional computer vision (CV) techniques and AI-based methods for obstacle detection. Camera-based visibility depends on the object’s visual contrast and color relative to the background. Figure~\ref{cameras} illustrates typical surveillance cameras at a level crossing and on-board visual sensors installed on a train.

Most surveillance cameras used at level crossings are RGB cameras, with CCTV systems being among the most common for traffic monitoring. Traditionally, much of the surveillance work has been performed by staff members \cite{yu2018railway}. To minimize the need for human oversight, numerous AI-based methods have been developed to enhance these surveillance systems \cite{yu2018railway}, \cite{zaman2019artificial}, \cite{wen2019general}. These methods typically involve annotating various types of intruders—such as humans, animals, and vehicles—and training neural networks for detection. This approach effectively transforms the problem of detecting foreign objects into a target detection problem. However, two main drawbacks of using a single camera with AI-based methods are: first, the system cannot detect objects that were not labeled during training; second, a single camera lacks the ability to provide stereo information about objects, which can lead to false alarms, such as mistaking a shadow for an obstacle \cite{ohta2005level}. 

To address these limitations, stereo cameras have been introduced at level crossings. A stereo camera system consists of two or more cameras placed at a fixed distance from each other, capturing images of the same object from slightly different angles. This creates a disparity between the images, which, along with the known distance between the cameras, allows for the calculation of the distance to the object. In \cite{ohta2005level}, the city block distance method is used to exclude sub-images that lack sufficient brightness differences, and obstacle detection is achieved by matching the current three-dimensional shape of an object with the background shape.

Most on-board video cameras are RGB cameras, which capture color images using three primary color channels: red, green, and blue. These cameras record image information through these three channels and then combine the light data to reproduce a broad spectrum of colors, resulting in a full-color image. RGB cameras are extensively used in the fields of computer vision and machine vision due to their ability to accurately capture and represent color information.

In obstacle detection using an on-board video camera, the first step is to detect the rail track, which serves as the region of interest (ROI) for subsequent obstacle detection. In most traditional computer vision methods, rail lane detection is approached as a line detection problem. A commonly used technique for this purpose is the Hough transform \cite{hough}, which identifies a line that best fits the majority of points within the region of interest \cite{rodriguez2012obstacle}, \cite{dynamic}. Other computer vision techniques for track detection also exist, as discussed in \cite{kudinov}, \cite{ross2010vision}, \cite{wang2018transportation}. Additionally, AI methods can be applied for track detection; for example, a Convolutional Neural Network (CNN) model is used in \cite{wang2018efficient} to detect rail tracks. This method consists of two parts: extraction of the rail region and subsequent refinement. The refinement process optimizes the alignment of the extracted rail region contours using a polygon fitting method, resulting in a more precise rail region contour. 

Once the track is detected, obstacle detection can be applied within the region of interest. In the obstacle detection phase described in \cite{rodriguez2012obstacle}, the process begins with the application of the Canny algorithm to detect edges \cite{rong2014improved}. Next, contours within the image are identified and emphasized, with small and isolated objects being removed. After the contours are filled in, a systematic search is conducted, guided by the rail tracks. Each rail is independently tracked using a small dynamic region that moves from the bottom to the top of the rail. During each analysis step, the region and its centroid are calculated. If these metrics deviate significantly from the expected values, the region is expanded, and its centroid is adjusted to encompass any abnormal objects~\cite{rodriguez2012obstacle}.

AI-based methods recently proposed for railway obstacle detection primarily utilize deep learning (DL) networks that have been validated in other scenarios. These DL networks are typically trained on large, publicly available object detection datasets or re-trained with custom-made railway datasets \cite{ristic2021review}. For example, in \cite{yu2018railway}, a DL-based method is presented for detecting obstacles on rail tracks, transforming the obstacle detection problem into a target object detection problem.

The visual sensors mentioned above provide detailed information about the level crossing and railway environment. However, their performance is highly dependent on visibility and can significantly decline in poor weather conditions or low-light environments. In particular, effective operation at night often requires additional illumination, such as dedicated lighting systems that could be activated when a train approaches a level crossing. To overcome visibility-related limitations, systems that integrate various types of cameras have been developed to enhance detection performance under all conditions. One such system is the SMART on-board multi-sensor obstacle detection system \cite{ristic2022smart}. This system combines RGB cameras, a night vision camera, and a thermal camera, utilizing a machine learning-based method for obstacle detection. This approach reduces the system's dependency on lighting and environmental conditions. However, a critical challenge in such multi-sensor systems is the "registration" process—aligning images from different cameras to ensure that corresponding pixels from each sensor map to the same geographic coordinates. 

Another novel vision-based sensor worth noting is the event camera. Unlike traditional cameras that capture images at fixed intervals, event cameras asynchronously measure per-pixel brightness changes and output a stream of events encoding the time, location, and sign of the brightness changes \cite{gallego2020event}. These sensors have already been applied to detect dynamic obstacles \cite{sanket2020evdodgenet, falanga2020dynamic} and enhance night vision obstacle detection \cite{yasin2020night}, making them a promising technology for deployment at level crossings or in on-board systems.

\subsection{Radar}
\label{radar_intro}
The performance of inductive loops is significantly limited by their detection range and the types of objects they can identify. Although various cameras can provide relatively
accurate obstacle detection, their effectiveness is greatly
reduced in rain, snow, or low-light conditions. To address
these limitations, radar technology has also been introduced
for railway obstacle detection.

Radar technology provides reliable environmental sensing capabilities due to its fundamental operating principle: it emits radio-frequency waves and detects their reflections from objects. {Depending on the system configuration, radar can be classified as monostatic, where the transmitter and receiver are co-located, or bistatic, where the transmitter and receiver are placed at separate locations. The choice between these configurations influences aspects such as angular coverage, synchronization requirements, and sensitivity to clutter.} Unlike passive sensors such as cameras, which rely on ambient light, radar is an active sensing system that generates its own illumination. This self-illumination and the use of longer wavelengths (typically in the millimeter to centimeter range) enable radar signals to penetrate visual obstructions such as fog, rain, and dust more effectively than optical signals. Physically, longer wavelengths are less susceptible to scattering by particles that are small relative to the wavelength, explaining radar's robustness under adverse environmental conditions.

However, these benefits come with trade-offs. Radar systems generally offer lower spatial resolution than optical sensors, which operate in the visible light spectrum with much shorter wavelengths. The angular resolution of radar is limited by both wavelength and aperture size; Achieving finer resolution requires larger antenna arrays or advanced digital beamforming. {By contrast, cameras operating in the visible spectrum can exploit the short wavelength of light, combined with features such as shape, color, and texture, to achieve much higher spatial detail.}

Microwave and ultra-wideband (UWB) radar systems operate within the frequency range of 300 MHz to 300 GHz, enabling robust obstacle detection under adverse weather conditions such as rain, snow, and fog. These systems remain unaffected by visibility, ambient light, or sunlight interference \cite{HORNE_Evaluation}. Common microwave radar types include Continuous Wave (CW) Doppler radar, primarily detecting moving objects, and Frequency Modulated Continuous Wave (FMCW) radar, which can distinguish stationary objects by transmitting frequency-swept electromagnetic waves. FMCW radar determines the object's range by analyzing frequency shifts in reflected signals, thus effectively detecting both stationary and moving targets \cite{barrick1973fm, stove1992linear, stove2004modern}.

Although FMCW radar is prevalent, especially in automotive applications due to its resilience against mutual interference, this advantage is less critical for stationary, trackside installations. In such contexts, classical pulse radar or alternative wideband waveforms can be advantageous due to their high range resolution and simpler signal processing requirements. Evaluating these alternative radar technologies may yield more efficient and cost-effective solutions for railway obstacle detection systems.

Radar detection faces two primary challenges: clutter and resolution, both of which are influenced by wavelength and aperture size. Clutter refers to unwanted echoes from environmental objects such as buildings, trees, or atmospheric particles, which can obscure or interfere with target detection. The severity of clutter depends on the radar’s wavelength, with shorter wavelengths being more susceptible to scattering from small objects and environmental noise, making target discrimination more difficult in high-clutter environments. Resolution, on the other hand, determines the radar’s ability to distinguish between closely spaced objects. It is affected by both wavelength and aperture size: shorter wavelengths improve range resolution, enabling finer discrimination between targets along the radar beam, while a larger aperture enhances angular resolution by narrowing the beamwidth, allowing for better separation of objects within the radar’s field of view. The trade-off between clutter and resolution underscores the importance of selecting appropriate radar parameters. While shorter wavelengths offer superior resolution, they are more susceptible to clutter, necessitating advanced signal processing techniques for clutter suppression. Conversely, longer wavelengths reduce clutter interference but may compromise resolution. Optimizing these parameters is essential for ensuring reliable target detection and minimizing environmental interference.

Microwave radar has been applied to foreign object intrusion detection in railway environments. In~\cite{yanwei2021research}, its performance was compared with that of computer vision (CV) methods. {Since radar measurements exhibit systematic transverse errors that increase with longitudinal distance, the authors first conducted offline calibration by repeatedly measuring fixed targets, deriving an error-correction function that was then applied online. A safety clearance area was constructed according to railway gauge standards, and moving objects within this area were filtered and tracked using a Kalman filter. The state vector included abscissa, transverse velocity, ordinate, and longitudinal velocity, while observations consisted of the abscissa, ordinate, and longitudinal velocity. Thus, the method goes beyond static scene analysis and explicitly performs real-time tracking of obstacles as they enter the detection zone.} Field tests demonstrated that radar achieves a higher detection rate—defined as the ability to successfully detect obstacles, particularly at longer distances—than cameras. However, radar typically exhibits lower detection accuracy, referring specifically to the precision in identifying and distinguishing individual objects and their exact positions. When two objects are in close proximity, radar may merge them into a single target, whereas computer vision methods rely on object features such as color and texture, enabling them to distinguish between multiple closely spaced objects more effectively. Despite this limitation, merging closely spaced objects does not significantly compromise obstacle detection effectiveness in practical railway scenarios, as ensuring a high detection rate is often more critical for maintaining safety.

Another example of radar-based level crossing obstruction detection is the use of multiple-input multiple-output (MIMO) radar \cite{narayanan2011railway}. In this setup, an MIMO FMCW radar is mounted at a height of 4 meters, employing an antenna configuration with sixteen transmitting and eight receiving elements, forming an equivalent phased array antenna with 128 elements. The system applies two stages of signal processing: digital beamforming and background clutter removal. 

Similarly, ultra-wideband (UWB) radar has also been explored for railway safety applications. In \cite{govoni2015uwb}, a multi-static UWB radar system was proposed for railway crossing surveillance. This system can detect and localize obstacles and estimate their volume, even under static conditions, using 3D imaging. To reduce ambiguity in 3D image generation, the Fixed Object Scanner (FOS) algorithm performs multiple scanning phases, where only a subset of nodes is active in each phase. A binary hypothesis test is then applied to each 3D pixel to improve obstacle detection accuracy. {In this study, the authors consider UWB signals in the 3--5 GHz band, corresponding to a bandwidth of about 2~GHz. This is significantly larger than the bandwidth typically available for conventional microwave radars in railway applications (tens to a few hundred MHz), thereby enabling centimeter-level range resolution and coarse 3D imaging capabilities.}

While radar cannot match the environmental detail available from cameras, it offers clear advantages in poor lighting and adverse weather, making it especially suitable for safety-critical applications. Its physical resilience, active sensing nature, and versatility across wavelengths make radar a promising core technology for future railway level crossing obstacle detection systems.

\subsection{LiDAR}
Although radar sensors demonstrate great robustness in varying light conditions and extreme weather, they can only detect obstacles with sufficient volume, potentially overlooking smaller objects such as rocks, baskets, jerrycans, or even children. To enhance the precision of obstacle detection systems at level crossings, 3D laser radar, or LiDAR (Light Detection and Ranging), has been introduced. 

The principle of measurement of the 3D laser is very similar to the radar system. A 3-D laser radar emits a laser pulse to an object, and measures the time that it takes for reflected laser to return to the radar (time-of-flight method) to acquire a distance to that object. A precise 3D map of the environment is generated by scanning the entire area of a level crossing in the horizontal and vertical directions \cite{bachman1979laser}, \cite{heritage2009principles}.

In this approach, LiDAR point cloud coordinates that are elevated above the road surface are first extracted based on their spatial values. Points that are spatially close are grouped into clusters, each representing a potential object. These clusters are then analyzed to estimate the positions and dimensions of the detected objects. By continuously performing these operations, the system can recognize and classify objects in the scene. Furthermore, by tracking changes in the positions of these clusters over time, the system is capable of estimating their speeds and directions of movement~\cite{hisamitsu20083}.

\begin{figure*}[t]
    \centering
    \begin{subfigure}[b]{0.32\textwidth}
        \includegraphics[width=\textwidth]{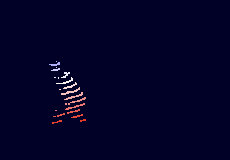}
        \caption{Pedestrian}
    \end{subfigure}
    \hfill
    \begin{subfigure}[b]{0.32\textwidth}
        \includegraphics[width=\textwidth]{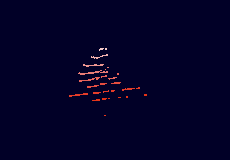}
        \caption{Cyclist}
    \end{subfigure}
    \hfill
    \begin{subfigure}[b]{0.32\textwidth}
        \includegraphics[width=\textwidth]{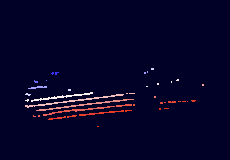}
        \caption{Car}
    \end{subfigure}
    \caption{Example LiDAR point clouds collected at a level crossing using a Velodyne VLP-16 sensor. Three types of obstacles are shown: (a) a pedestrian, (b) a cyclist, and (c) a car crossing the tracks.}
    \label{Lidar_measure}
\end{figure*}

Figure~\ref{Lidar_measure} illustrates three LiDAR point cloud frames captured using a Velodyne VLP-16 sensor~\cite{ouster_vlp16_2025} installed at a level crossing. The frames depict typical scenarios involving different types of moving objects: (a) a walking pedestrian, (b) a person on a bicycle, and (c) a car. {These examples demonstrate the types of obstacles encountered in level crossing environments and highlight the potential suitability of LiDAR sensors for obstacle detection tasks.}

A 3D laser radar manufactured by IHI is showcased in \cite{hisamitsu20083}, though it is primarily designed to detect obstacles with a volume greater than $\mathrm{1 m^3}$. In \cite{amaral2016laser}, an improved method is proposed to enhance the detection of relatively smaller objects, down to volumes above $10  \mathrm{dm^3}$. Obstacle detection is based on the assumption that the ground plane at a level crossing is quasiplanar and dominant. A background model of the crossing environment is constructed from a series of point cloud frames acquired during an offline training phase when no obstacles are present. This model, regularly updated to adapt to environmental changes and maintain accuracy, enables the system to distinguish between the typical static background and new, potentially hazardous objects. Obstacles are identified as objects appearing in real-time point clouds that differ from the background model. To streamline deployment, the method includes an automatic calibration process, which estimates the spatial position (pose) of both the level crossing and railway relative to the laser scanner, eliminating the need for manual on-site calibration.

\subsection*{Summary}
Each sensing modality for obstacle detection at level crossings presents distinct advantages and limitations. Inductive loops are weather-resilient but limited to metallic objects. Cameras offer rich semantic information but degrade in low visibility. Radar provides robustness under adverse weather yet suffers from lower resolution. LiDAR enables precise 3D mapping, effectively detecting various object types, but its performance may deteriorate in heavy rain or snow. The  comparison of sensor technologies for obstacle detection at railway level crossings is shown in Table \ref{tab:sensor_comparison}.

{Beyond these sensor-specific characteristics, a common processing requirement across all modalities is the ability to identify deviations from a reference state where no obstacle is present. This typically involves background modeling or subtraction techniques, which serve as the foundation for distinguishing newly introduced objects in the monitored area. Depending on the sensing modality, this may be implemented through statistical background estimation in inductive loops, pixel- or feature-based subtraction in cameras, clutter suppression in radar, or point cloud differencing in LiDAR.}

\begin{table*}[htbp]
\renewcommand{\arraystretch}{0.95}
\scriptsize

\caption{Comparison of Sensor Technologies for Obstacle Detection at Railway Level Crossings}
\centering
\begin{tabular}{|l|p{2.8cm}|p{2.8cm}|p{2.4cm}|p{2.4cm}|}
\hline
\textbf{Sensor Type} & \textbf{Advantages} & \textbf{Limitations} & \textbf{Environmental Robustness} & \textbf{Detection Capabilities} \\
\hline
Inductive Loops & 
- Reliable in adverse weather \newline 
- Long lifespan \newline 
- Mature technology &
- Detects only metallic objects \newline 
- Requires intrusive installation \newline 
- Limited spatial coverage &
High (resistant to rain, fog, snow) &
Vehicles and trains with metallic mass \\
\hline
Cameras &
- Rich semantic info \newline 
- Well-developed AI methods \newline 
- Affordable installation &
- Sensitive to lighting and weather \newline 
- Limited depth perception (monocular) &
Low (susceptible to glare, rain, fog) &
Humans, vehicles, animals (with training) \\
\hline
Stereo Cameras &
- Depth estimation \newline 
- Improved obstacle localization &
- Requires calibration \newline 
- Still affected by visibility issues &
Medium (suffers in fog/rain) &
3D object localization, shape estimation \\
\hline
Thermal/Night Vision Cameras &
- Detects heat sources in low light \newline 
- Operates in darkness &
- Limited object discrimination \newline 
- Challenging calibration in multi-sensor setups &
Medium–High (robust in darkness, fog) &
Human and animal detection at night \\
\hline
Radar (CW/FMCW/UWB) &
- Robust in all weather/light conditions \newline 
- Active sensing \newline 
- Long detection range &
- Low spatial resolution \newline 
- Struggles with close object separation &
Very High (penetrates fog, rain, dust) &
Moving and stationary object detection \newline Basic classification \\
\hline
LiDAR &
- High spatial resolution \newline 
- Accurate 3D mapping \newline 
- Works in daylight and darkness &
- Sensitive to weather (rain, fog) \newline 
- Expensive and power-consuming &
Medium (affected by rain and dense fog) &
Fine-grained obstacle detection, object tracking \\
\hline
\end{tabular}
\label{tab:sensor_comparison}
\end{table*}

\section{Impact of the adverse weather conditions}

Extreme weather conditions, including heavy rain, snow, fog, hail, etc., can substantially degrade sensor detection performance by introducing noise, distortion, or attenuation to the collected data. To effectively address these challenges, it is crucial to first establish a systematic understanding of how adverse weather impacts sensor data acquisition. Certain sensors, such as inductive loops, are minimally affected by harsh weather, while others, such as radar sensors, exhibit moderate resilience. In contrast, sensors such as cameras and LiDAR are highly susceptible to weather-induced degradation.

This section offers a comprehensive overview of the effects of adverse weather conditions on various sensors. The subsequent section will explore existing methodologies developed to mitigate these impacts.
\subsection{Cameras}
\label{adverse_camera}
While cameras perform reliably under clear conditions, their effectiveness degrades significantly in adverse weather. The impact of extreme weather on camera sensors can be broadly categorized into two types. The first includes dynamic conditions such as rain, snow, and hail, which introduce complex effects including intensity fluctuations, contrast changes, and occlusions. The second involves more stable phenomena like fog and haze, which primarily reduce image contrast. This section will discuss these two categories of weather effects separately.

\subsubsection{Rain, Snow and Hail}
Adverse weather conditions, such as rain, snow and hail, can introduce significant intensity fluctuations in captured images and videos. These variations arise from changes in lighting, reflections, and occlusions caused by precipitation. Most algorithms designed for outdoor vision systems assume that the intensity of image pixels is directly proportional to the brightness of the scene. However, adverse weather violates this assumption, resulting in degraded image quality and reduced system performance \cite{AdverseWeather}, \cite{CameraSeeRain}. For instance, falling raindrops can scatter and refract light before it reaches the camera sensor, leading to localized dimming and distortion of background features. This interaction often results in blurred or smeared regions in the image, particularly along object edges. Similarly, heavy snow and hail can fluctuate image intensity and obscure the edges of objects, making them difficult or impossible to recognize in the image or video \cite{AdverseWeather}, \cite{DetectRain}.

In addition to their impact on image quality, raindrops, snowflakes, and hailstones can directly affect the functionality of the camera system itself. For instance, low temperatures may cause optical and mechanical disruptions that impair the camera's operation. During a hailstorm, the camera lens is vulnerable to physical damage. Rainy conditions can lead to electrical and optical malfunctions; if the camera system lacks adequate waterproofing, it is at risk of short circuits caused by water ingress. Optically, raindrops on the lens can alter the camera's focus, resulting in parts of the captured image appearing blurred and out of focus~\cite{AdverseWeather,nava2019weather}.

Moreover, snow and ice can accumulate on the lens surface, obstructing the field of view entirely and rendering the camera temporarily blind. These disruptions compromise the effectiveness of image processing tasks such as pattern recognition, object tracking, or detection, and can ultimately lead to partial or complete system failure.

\subsubsection{Fog and Haze}
Fog and haze degrade image quality primarily by reducing image contrast, which increases the difficulty of recognizing pattern edges. This degradation arises from the scattering and absorption of light by a high concentration of atmospheric aerosols and particles. In addition to contrast loss, fog and haze can produce halo-like artifacts around bright light sources—such as street lamps or vehicle headlights—due to forward scattering. Back-scattering of light toward the camera sensor is also common, particularly in night-time or low-light scenarios, and can result in veiling glare that further reduces image clarity~\cite{miclea2021visibility}.

Furthermore, increased moisture and condensation associated with fog can accumulate on the camera lens, obscuring the optical path. This leads to additional blurring, shadowing, and distortion in the captured image, further impairing the performance of visual detection systems~\cite{AdverseWeather,hazeremoval}.

\subsection{Radar}
\label{radar_weather}
The impact of adverse weather on radar sensors can be categorized into two primary effects: attenuation and backscattering. Attenuation reduces the received signal power due to energy loss as the electromagnetic wave propagates through a medium, while backscattering introduces interference by reflecting signals from dispersed particles back toward the radar receiver. Various weather conditions, including rain, fog, snow, hail, dust, and sand, influence radar detection in broadly similar ways, differing primarily in their particle size distributions, concentrations, and dielectric properties.

Research conducted at the Ballistic Research Laboratory in 1988 systematically measured the attenuation and backscattering effects of different weather phenomena. Among them, rain was found to have the most significant impact on radar performance, primarily because the size of raindrops is often comparable to the millimeter wave radar wavelength~\cite{Ballistic}.

Unlike optical systems such as cameras, radar is generally less affected by diffraction effects caused by sub-wavelength aerosols (e.g., fog, haze, or dust), especially at centimeter wavelengths, since these particles are usually much smaller than the radar wavelength and do not produce substantial diffraction-based interference. However, at millimeter-wave bands, hydrometeors such as raindrops can be comparable to the wavelength, leading to significant Mie scattering and rain attenuation~\cite{Ballistic,pozhidaev2010estimation}. In contrast, cameras, operating at visible-light wavelengths (hundreds of nanometers), are highly susceptible to diffraction in the presence of fine particles and also suffer from optical phenomena such as glare, caustics, or lens-based aberrations under adverse weather. These differences reinforce radar’s robustness in poor visibility environments, while also highlighting its relatively lower spatial resolution and detail fidelity compared to optical systems. Furthermore, the mathematical models for attenuation and backscatter in snow and mist are analogous to those used for rain, and their detailed effects are provided in the referenced works~\cite{pozhidaev2010estimation}.

\subsubsection{Rain Attenuation Effect}
The attenuation effect of rain on millimeter-wave radar can be modeled using a probabilistic expression for the received signal power, as presented in~\cite{AdverseWeather}:

\begin{equation}
P_r = \frac{P_t G^2 \lambda^2 \sigma_t}{(4 \pi)^3 r^4} \cdot V^4 \exp(-2 \gamma r),
\end{equation}
where $P_r$ and $P_t$ denote the received and transmitted power, respectively; $\lambda$ is the wavelength; $G$ is the antenna gain; $\sigma_t$ is the radar cross-section (RCS) of the target; and $r$ is the distance between the radar and the target. The term $V$ represents the multipath coefficient, while $\gamma$ is the rain attenuation coefficient, which depends on the rainfall rate.

For further interpretability, the expression can be converted into the dB scale:
\begin{equation}
\begin{split}
P_{r,\mathrm{dB}}
  = P_{t,\mathrm{dB}}
    + 2\,G_{\mathrm{dB}}
    + 2\,\lambda_{\mathrm{dB}}
    + \sigma_{t,\mathrm{dB}}
    - 30\log_{10}(4\pi) \\
    - 40\log_{10} r 
    + 4\,V_{\mathrm{dB}}
    - 8.686\,\gamma\,r.       
\end{split}
\end{equation}

From this equation, it is evident that the received signal power is influenced by path loss, multipath propagation, and rain-induced attenuation. Path loss refers to the reduction in signal strength as the wave propagates through space and increases with distance due to geometric spreading and absorption. Multipath propagation occurs when signals reach the receiver via multiple paths caused by reflection, diffraction, and scattering from surrounding objects such as buildings. This results in constructive and destructive interference, as well as phase shifts. The multipath effect can be analytically modeled using the formulation provided in~\cite[Eq. 4]{kulemin1999influence}. Rain attenuation is determined by the attenuation coefficient $\gamma$, which depends on the rain rate, wave frequency, and raindrop-size distribution and can be computed using empirical models provided in\cite{pozhidaev2010estimation}.

\subsubsection{Rain Backscattering Effect}
The rain backscattering effect increases self interference at the receiver, necessitating the maintenance of a minimum threshold for the signal-to-interference-plus-noise ratio (SINR) to ensure accurate obstacle detection. The relationship between the power intensity of the target signal ($S_t$) and the backscatter signal ($S_b$) is expressed as:

\begin{equation}
\frac{S_t}{S_b} = \frac{8 \sigma_t}{\tau c \theta_{BW}^2 \pi r^2 \sigma_i}
\end{equation}
where $c$ is the speed of light, $\theta_{BW}$ is the antenna beamwidth, $\tau$ is the pulse duration, and $\sigma_i$ is the rain backscatter coefficient. The rain backscatter coefficient $\sigma_i$ depends on the rain rate, wave frequency, and raindrop-size distribution which can be calculated using the methodologies described in \cite{pozhidaev2010estimation} and \cite{huang2001rain}.

In addition to the attenuation and scattering models, a physical propagation model for radar in random media (such as rain) was proposed in \cite{yektakhah2024model}, \cite{ibrahim2014simulation}, and \cite{yektakhah2023physics}, which specifically models phase distortion caused by multiple scattering from randomly distributed particles. In these studies, an $S$-matrix method, which is a general approach applicable to both dense and sparse random media, was applied to rain media. The $S$-matrix method divides the random medium into multiple slabs of a certain thickness. Each slab is modeled as a network with multiple input and output ports, where the ports represent rays with different polarizations propagating in various directions as they enter and leave the slabs. Since this model is capable of preserving phase information, it holds promise for beamforming applications, enabling the development of new algorithms to mitigate the impact of adverse weather conditions on radar performance. Beyond rain media, this modeling method can also be applied to other weather conditions, such as foliage, sandstorms, and cloud or fog \cite{yektakhah2023physics}.

Following the methodology in~\cite{yektakhah2024model}, for two points located on the same wavefront at a range $R$ from the source and separated by a distance $d$, the rain-induced phase shift and amplitude ratio are characterized by the probability density functions (pdfs) given in Equations~(9a) and (9b) of~\cite{yektakhah2024model}.

\begin{figure}[htbp]
\centerline{\includegraphics[width=\columnwidth]{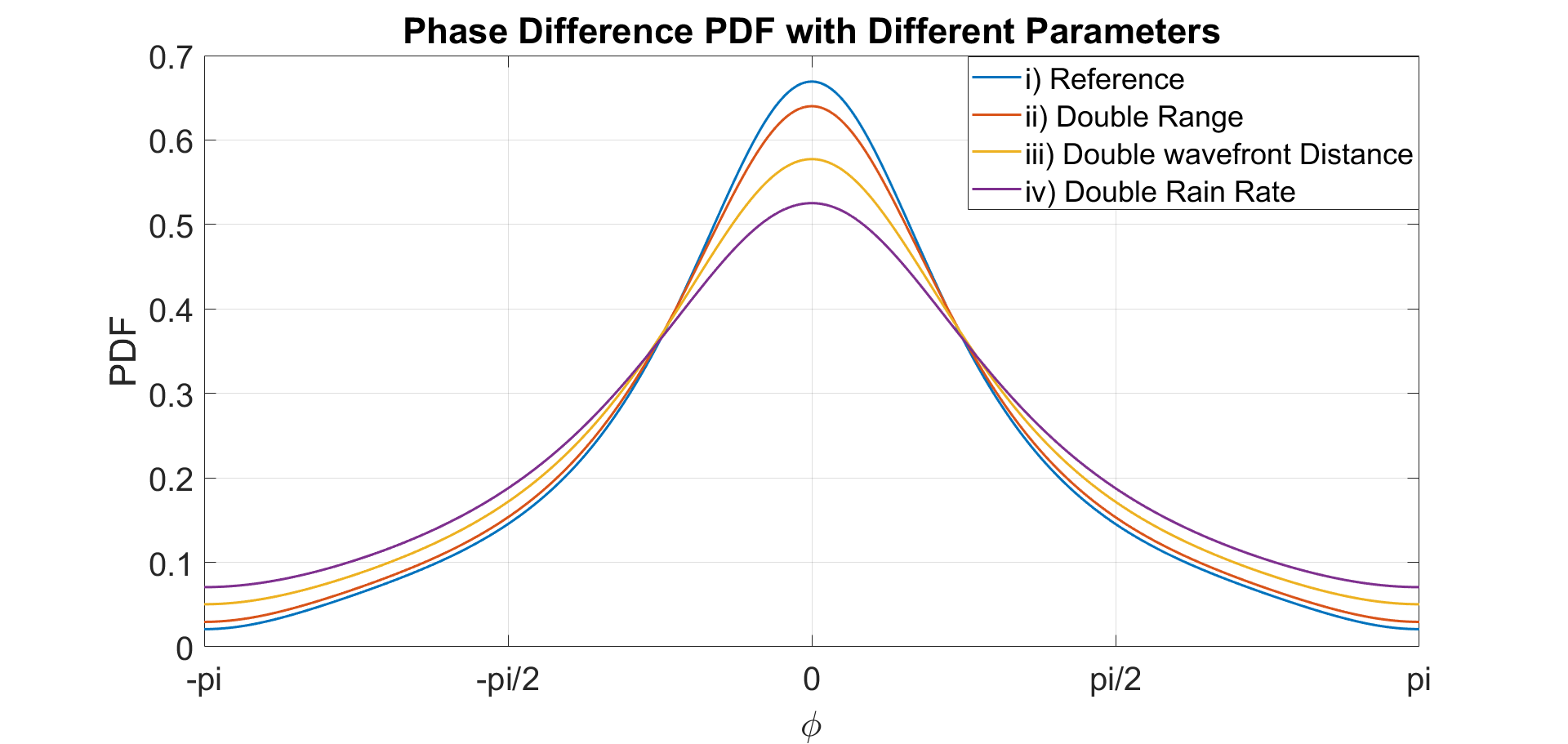}}
\caption{Phase difference pdf under different propagation scenarios.}
\label{pdf_demo}
\end{figure}

Figures~\ref{pdf_demo} illustrate the pdfs of the phase difference under varying propagation conditions. Four scenarios are evaluated, with corresponding parameter settings and computed $\alpha$ values summarized in Table~\ref{parameter_table}. The reference case (i) uses $d = 4 \lambda_0$, $R = 200\ \mathrm{m}$, and a rain rate of $25\ \mathrm{mm/hr}$. In case (ii), the range is doubled to $400\ \mathrm{m}$; in case (iii), the wavefront separation is doubled to $8 \lambda_0$; and in case (iv), the rain rate is increased to $50\ \mathrm{mm/hr}$, while other parameters remain unchanged.

\begin{table}[htbp]
\centering
\caption{Parameter settings and corresponding $\alpha$ values for Figures~\ref{pdf_demo}}
\begin{tabular}{c|c|c|c|c}
\hline
Case & $d$ $(\lambda_0)$ & $R$ (m) & Rain Rate (mm/hr) & $\alpha$ \\ 
\hline
(i) & 4 & 200 & 25 & 0.6470 \\[2pt]
(ii) & 4 & 400 & 25 & 0.6217 \\[2pt]
(iii) & 8 & 200 & 25 & 0.5598 \\[2pt]
(iv) & 4 & 200 & 50 & 0.4994 \\[2pt]
\hline
\end{tabular}
\label{parameter_table}
\end{table}

\subsection{LiDAR}
\label{lidar_weather}
Similar to cameras, LiDAR is also severely affected by adverse weather conditions. In addition to equipment malfunctions caused by sensor freezing or other mechanical complications under low temperatures, adverse weather impacts the intensity and point cloud features through backscattering or diversion when encountering particles such as raindrops, snowflakes, and others within the medium, making detection more challenging~\cite{dreissig2023survey}.

Another important consideration is that many LiDAR systems employ mechanically rotating parts to scan the environment in different directions. These moving components can be prone to wear and mechanical failure, especially under harsh environmental conditions such as extreme cold, where lubrication performance may degrade and materials may become brittle. This raises concerns about long-term reliability and maintenance requirements in real-world deployments. As an alternative, solid-state LiDAR systems—designed with no moving parts—offer the promise of increased durability and potentially higher reliability \cite{li2022progress}. However, as this technology is still relatively new, comprehensive data on its long-term performance and reliability under adverse weather conditions remains limited.

Theoretically, the influence of adverse weather such as rain, fog, or snow is analyzed in~\cite{rasshofer2011influences}, where the scattering from precipitation particles is modeled as the interaction of electromagnetic waves with dielectric spheres, using Mie theory as described in~\cite{bohren2008absorption}.

\section{Mitigation Techniques for Adverse Weather Effects}

To mitigate adverse weather effects at level crossings, two main strategies can be adopted. The first focuses on developing enhancement algorithms tailored to each individual sensor, while the second leverages sensor fusion by integrating data from multiple sensors to enhance the overall system robustness.

As an additional consideration, it is also possible to implement monitoring systems that assess environmental conditions in real time. When weather conditions are detected to be too poor for reliable obstacle detection, the railway system can switch to a "functionally degraded mode," wherein trains operate at reduced speed to maintain safety.

For single-sensor mitigation, the objective is to exploit the statistical characteristics of different weather conditions or to identify patterns in weather-induced distortions, enabling the recovery or estimation of clean data. In contrast, sensor fusion strategies provide redundancy, allowing the system to continue functioning reliably even if one or more sensors fail under adverse conditions. Furthermore, combining data from heterogeneous sensors allows for cross-validation, improving detection accuracy and reliability.

Sensor fusion can be implemented under different architectural paradigms. In \emph{early fusion}, raw sensor data are integrated at the initial stage of processing, while in \emph{late fusion}, each sensor's data is processed independently before being merged at a higher decision level \cite{wang2024fusion}. 

The following sections discuss and compare representative enhancement algorithms for individual sensors under adverse weather conditions, along with sensor fusion strategies aimed at improving system resilience and detection performance.

\subsection{Cameras}
As discussed in~\ref{adverse_camera}, the quality of visual sensor images can be significantly degraded under adverse weather conditions, and in some cases, the sensor’s functionality may be directly impaired. In this section, we focus specifically on the degradation of image quality. Issues related to sensor failure can be addressed through sensor fusion-ensuring operational continuity when certain sensors malfunction—or through physical protection measures, such as camera covers or integrated heating devices.

\subsubsection{Fog and Haze}

Fog and haze, as relatively stable weather phenomena, primarily degrade image contrast. Numerous image dehazing algorithms have been developed to address this issue. In~\cite{he2010single}, He \emph{et al.} proposed a single-image haze removal method based on the dark channel prior, which exploits the observation that in most local patches of outdoor haze-free images, at least one color channel exhibits low intensity. This prior enables accurate estimation of transmission maps and atmospheric light, allowing haze removal. A soft matting algorithm is used to refine the transmission map, reducing artifacts and improving visual quality. The method also produces a depth map as a byproduct. However, its performance degrades in scenes with light-colored surfaces or wavelength-dependent scattering, which can lead to errors in transmission estimation and color fidelity.

Similarly, Zhu \emph{et al.}~\cite{zhu2015fast} proposed a fast dehazing algorithm based on a color attenuation prior, which models the relationship between haze concentration, depth, brightness, and saturation. A linear model is trained to estimate scene depth, from which the transmission map is derived and used for image restoration. This method is claimed to achieve high computational efficiency and performs well in challenging regions such as white or gray surfaces. Nonetheless, its reliance on supervised learning makes it sensitive to training data quality and less robust in highly variable atmospheric conditions.

Except for these prior-based methods, there are also deep learning-based methods for the haze-removal problem. One of the most famous algorithms is DehazeNet \cite{cai2016dehazenet}, which was among the first to use convolutional neural networks (CNNs) for haze removal. DehazeNet directly estimates the transmission map from hazy images by leveraging a carefully designed CNN architecture, significantly improving computational efficiency and performance compared to traditional methods. Following DehazeNet, several advanced deep learning frameworks have emerged, including AOD-Net \cite{li2017aodnet}, which integrates atmospheric scattering models directly into the network, and FFA-Net \cite{qin2020ffa}, which employs attention mechanisms to enhance feature extraction and restore visibility. These methods  demonstrate that deep learning approaches can effectively model the complex characteristics of haze, thus providing superior haze removal performance.

\subsubsection{Rain, Snow, and Hail}

Rain, snow, and hail are dynamic weather phenomena that introduce temporally and spatially varying structured noise in images, making their removal particularly challenging. Unlike steady degradations such as fog, these conditions often result in complex distortions that require more sophisticated modeling techniques.

Traditional approaches based on dictionary learning and sparse coding~\cite{kang2012automatic} have been largely replaced by data-driven methods. Recent single-image deraining models such as DDN~\cite{fu2017removing} and RESCAN~\cite{li2018recurrent} employ deep convolutional neural networks (CNNs) to progressively isolate and remove rain streaks. For video-based tasks, temporal information is further leveraged to enhance robustness, as shown in~\cite{yang2019frame}, which exploits motion consistency to better separate rain artifacts from the background.

In~\cite{Lionline2021}, Li et al. propose an online multi-scale convolutional sparse coding (OMS-CSC) framework for rain and snow removal in surveillance videos. The method employs temporally adaptive parameters and an affine transformation operator to address noise induced by weather conditions and background motion caused by camera jitter, scaling, and distortion. Experimental evaluations on synthetic and real-world datasets report improvements in visual quality and quantitative metrics, suggesting the approach's applicability in real-time scenarios.

In~\cite{li2024dual}, Li et al. propose a raindrop removal method using dual-pixel (DP) sensor data, named Dual-Pixel Raindrop Removal Network (DPRRN). This approach utilizes the disparity between left and right sub-images captured by a DP sensor, especially prominent in out-of-focus raindrops, facilitating their detection and removal. The DPRRN method consists of two stages: DP-based raindrop detection and DP-guided image restoration. Evaluations on synthetic and real-world datasets demonstrate its effectiveness, showing improved performance compared to several existing methods.

\subsection{Radar}

As discussed in Sections~\ref{radar_intro} and~\ref{radar_weather}, radar sensors are relatively robust under adverse weather conditions. The primary challenges affecting their performance are attenuation and backscattering. Due to the limited availability of models accurately describing wave propagation in adverse weather, few methods have been specifically proposed to enhance radar sensing performance under such conditions—aside from sensor fusion approaches. A notable advancement is the statistical model for wave propagation in rain presented in~\cite{yektakhah2024model}. Building on this model, a recent development in~\cite{yan2025robust} introduces a covariance-based Direction-of-Arrival (DoA) estimation method tailored for rain-induced distortion.

\subsubsection{A Model-based Method}
The approach presented in~\cite{yan2025robust} models the electric field fluctuations across antenna array elements as zero-mean, circularly symmetric complex Gaussian variables. These fluctuations are characterized by a Hermitian Toeplitz covariance matrix, leveraging its structured correlation among array elements. Utilizing this property significantly reduces the number of parameters needed for covariance matrix estimation, thereby simplifying the calibration process.

\begin{figure}[htbp]
\centerline{\includegraphics[width=\columnwidth]{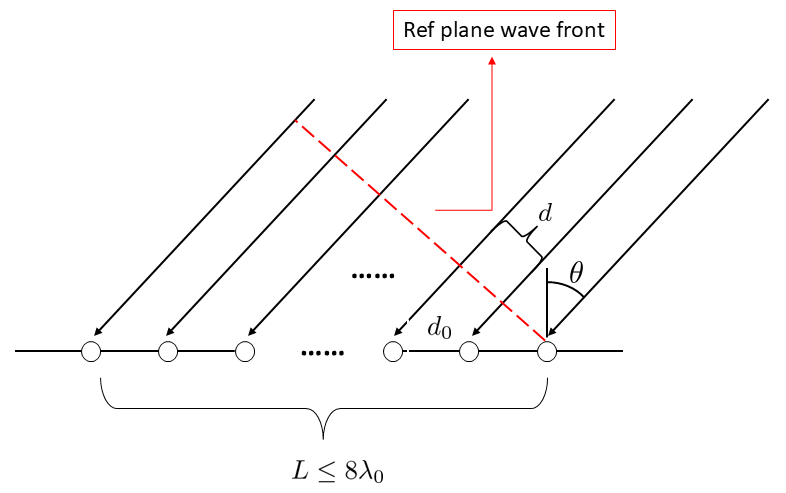}}
\caption{Illustration of the array measurement model with the rain distortion.}
\label{array_model}
\end{figure}

A uniform linear array (ULA) comprising $M$ antennas with spacing $d_0$ between adjacent elements is illustrated in Figure~\ref{array_model}. We assume electric field fluctuations occur across the wavefront upon reaching the first antenna during propagation; this wavefront is termed the \emph{reference plane wavefront}. Additionally, fluctuations beyond this reference plane are considered negligible.

Specifically, the received signal vector at a ULA under rain-induced distortion is modeled as:
\begin{equation}
\mathbf{y}(t) = [\mathbf{a}(\theta)\odot \mathbf{b}(t)] s(t) + \mathbf{n}(t),
\end{equation}
where $\mathbf{a}(\theta)$ represents the array steering vector, $\mathbf{b}(t)$ encapsulates the rain-induced distortions, $s(t)$ is the transmitted signal, and $\mathbf{n}(t)$ represents additive noise.

The covariance matrix of the received signals is then expressed as:
\begin{equation}
\begin{split}
\mathbf{R}_y &= E \{ \mathbf{y}(t) \mathbf{y}(t)^H \} \\
&= \underbrace{\mathbf{a}(\theta) \sigma_s^2 \mathbf{a}(\theta)^H}_{\mathbf{R}_x(\theta)} \odot \underbrace{E \{\mathbf{b}(t) \mathbf{b}(t)^H \}}_{\mathbf{R}_b} + \mathbf{R}_n,
\label{covariance measurement}
\end{split}
\end{equation}
where $R_b$ is the covariance matrix of the distortion vector, structured as a real-valued symmetric Toeplitz matrix derived from the rain-induced statistics.

A generalized least squares (GLS) covariance matching technique ~\cite{OTTERSTEN1998185} is employed to calibrate and recover the undistorted covariance matrix. This approach solves the optimization problem:
seek to minimize the following cost function:
\begin{equation}
(\hat{\theta},\hat{\mathbf{R}}_b)
= \arg\min_{\theta,\mathbf{R}_b} 
\left\lVert 
\mathbf{\hat{R}}_{y} \;-\;  \mathbf{R}_x(\theta) \odot \mathbf{R}_b
\right\rVert_{F}^{2}. 
\label{cost func}
\end{equation}
where $\hat{R}_y$ is the estimated covariance from measurements.

\begin{figure}[htbp]
    \centering
    \includegraphics[width=\columnwidth]{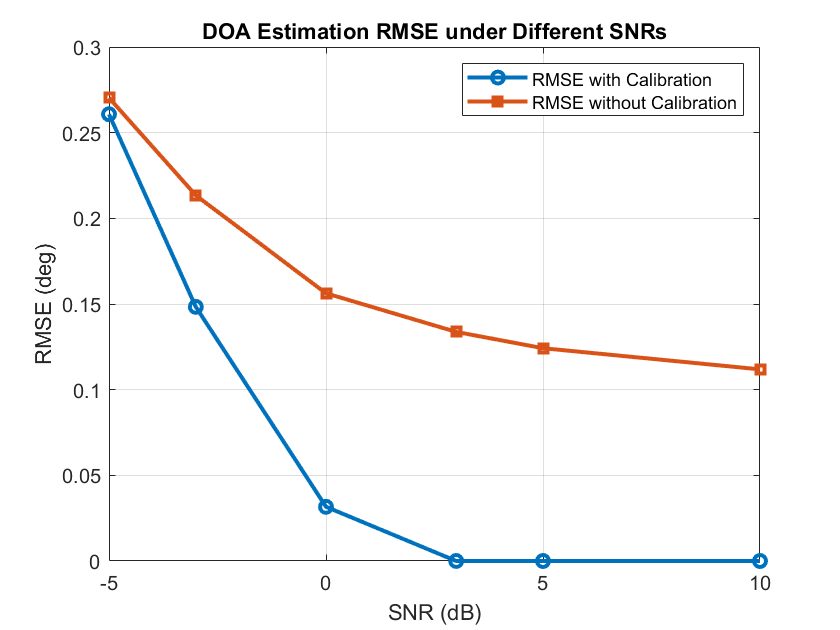}
    \caption{RMSE comparison between calibrated root-MUSIC-based DoA estimation and conventional root-MUSIC without calibration.}
    \label{rmse}
\end{figure}

Simulation results demonstrate that the proposed GLS-based calibration significantly enhances DoA estimation accuracy under adverse weather conditions. Specifically, it achieves a notable improvement in root mean squared error (RMSE) compared to conventional methods, as illustrated in Figure~\ref{rmse}, confirming its effectiveness in mitigating weather-induced distortions.

\begin{figure}[htbp]
    \centering
    \begin{subfigure}{\columnwidth}
        \centering
        \includegraphics[width=\linewidth]{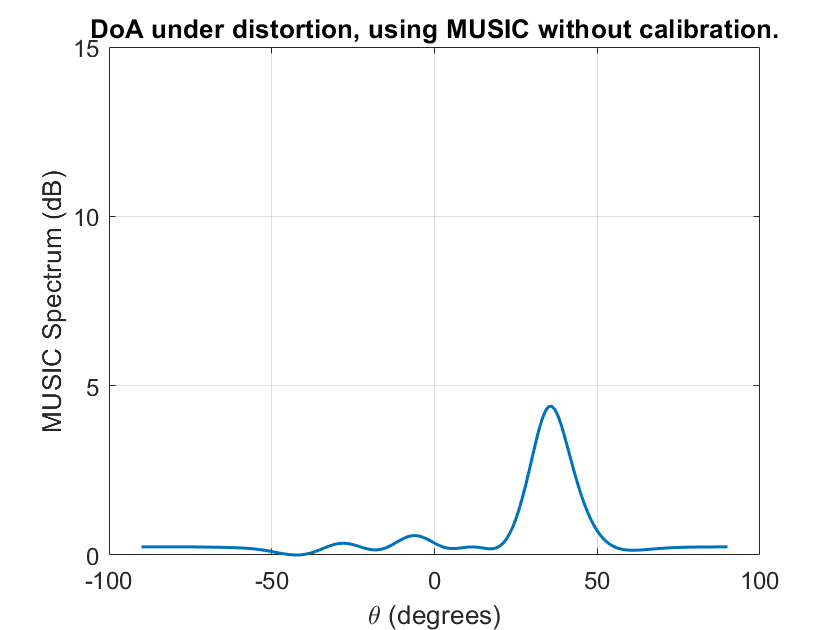}
        \caption{Rain, no calibration}
        \label{fig:MUSIC_rain}
    \end{subfigure}
    \hfill
    \begin{subfigure}{\columnwidth}
        \centering
        \includegraphics[width=\linewidth]{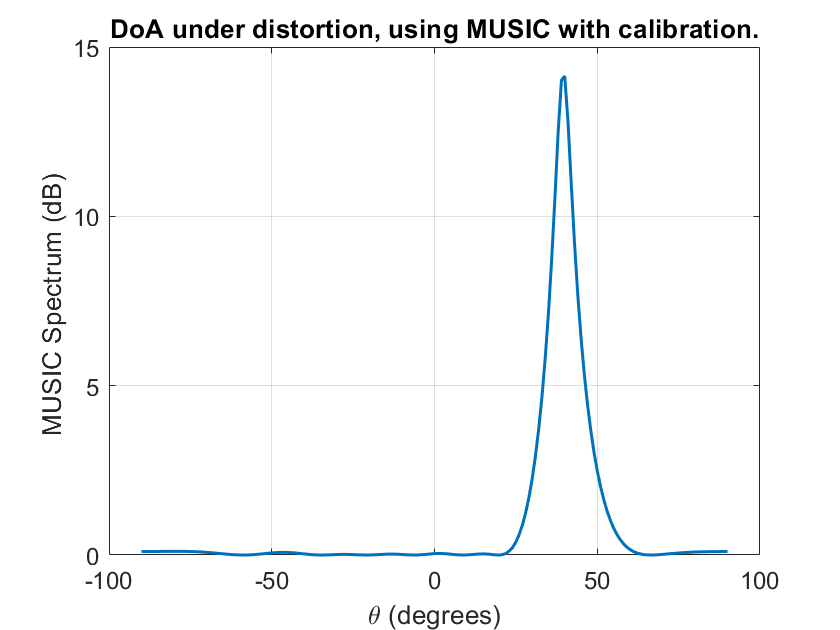}
        \caption{Rain, with calibration}
        \label{fig:MUSIC_calibration}
    \end{subfigure}
    \caption{MUSIC spectra in dB scale under two conditions: (a) rain distortion without calibration, and (b) rain distortion with calibration.}
    \label{fig:MUSIC_all}
\end{figure}

Furthermore, numerical simulations shown in Figure~\ref{fig:MUSIC_all} highlight the method’s robustness by comparing the MUSIC spectra under two scenarios: (i) with distortion but without calibration, and (ii) with both distortion and calibration. The calibrated case exhibits substantially improved peak sharpness and localization accuracy, closely approximating the undistorted reference, thereby validating the effectiveness of the proposed calibration strategy.

Future research into extending this method to multiple-source scenarios would be valuable for further enhancing radar sensing performance.

% 图像部分提前

\subsubsection{Learning-based Methods}
Few studies have specifically addressed radar performance enhancement under adverse weather conditions, primarily due to two reasons. First, radar is often employed as an auxiliary sensor alongside visual systems for obstacle detection tasks. Second, there is a scarcity of radar datasets collected under diverse adverse weather scenarios. A notable dataset addressing this gap is presented in~\cite{RADIATE}, which provides approximately three hours of annotated radar imagery captured from a vehicle-mounted sensor, comprising over 200{,}000 labeled instances of road users. On average, each radar frame contains about 4.6 labeled objects. The dataset encompasses eight distinct object categories captured under various weather conditions (e.g., clear, night, rain, fog, and snow) and driving environments (e.g., parked, urban, motorway, suburban), thereby providing a valuable resource for developing and evaluating robust radar-based deep learning algorithms.

\subsection{LiDAR}
As mentioned in Section~\ref{lidar_weather}, LiDAR point cloud data suffers from noise augmentation, point dropout, and intensity shifts due to signal attenuation and backscattering under adverse weather conditions. In this section, we discuss current solutions aimed at mitigating these weather-induced effects on LiDAR sensors.

\subsubsection{Full‐Waveform LiDAR}
Full-Waveform LiDAR (FWL) systems demonstrate superior performance in challenging meteorological conditions by recording and analyzing the complete return waveform ~\cite{mallet2009full, wallace2020full}. This capability allows FWL to distinguish atmospheric backscatter from rain, fog, or snow from genuine surface reflections. Additionally, their sub-centimeter range resolution and sensitivity to weak returns enable accurate depth profiling, even under significant signal attenuation. While these advantages make FWL attractive for adverse-weather applications, the technology is still less mature than conventional discrete-return LiDAR, with current use mainly limited to airborne surveys and research prototypes rather than large-scale commercial deployment~\cite{wallace2020full}.

The Full-Waveform LiDAR (FWL) return can be modeled as a temporal pulse, represented as a non-normalized statistical mixture of single-surface returns~\cite{pellegrini2000laser}:
\begin{equation}
    F(i; k, \phi) = \sum_{j=1}^{k} f_{\mathrm{system}}(i; \beta_j, t_{0_j}) + B,
\label{fwl_eq}
\end{equation}
where $k$ is the number of returns, $\beta_j$ is the amplitude of the $j$-th return, $t_{0_j}$ is its peak position, and $B$ denotes the background noise level. 

Extracting waveform parameters as described in~\eqref{fwl_eq} is essential for interpreting complex scenes with FWL. Unlike conventional LiDAR systems, such as the Velodyne scanner~\cite{lidar2016hdl}, which typically assume one or two reflections from relatively flat surfaces with limited depth variation, FWL systems can detect multiple echoes. These echoes may arise from secondary reflections, transmission through transparent materials, or interactions with detailed surface structures (e.g., vegetation) whose scale is smaller than the laser beam width~\cite{wallace2013design}. In practical settings, such as automotive or level-crossing obstacle detection, this capability becomes especially important in the presence of obscuring media, such as rain or snow, where early backscattering from suspended particles may dominate before the signal reaches the intended targets.

To extract multiple returns from photon histograms, various algorithms have been proposed. One category involves Reversible Jump Markov Chain Monte Carlo (RJMCMC) methods~\cite{marin2007bayesian, tachella2019bayesian}, capable of detecting weak and closely spaced returns (down to 1~cm at 330~m~\cite{tachella2019bayesian}) but at high computational cost. Alternatively, convex optimization methods with sparsity-promoting regularization~\cite{shin2016computational, halimi2017restoration} model the photon counts via Poisson statistics. These frameworks often integrate spatial priors—such as smoothness in the return count $k$~\cite{hernandez2008multilayered}, consistency in depth and reflectivity~\cite{rapp2017few,lindell2018single}, or non-local spatial dependencies~\cite{marais2017proximal, halimi2019robust,chen2019learning}, to further enhance performance.

However, FWL systems have several limitations. The capture and digitization of complete waveforms at high sampling rates generate large volumes of data, imposing significant demands on storage, communication bandwidth, and real-time processing resources. These factors increase system complexity, size, weight, power consumption, and cost, making FWL systems less suitable for compact, low-power applications. Moreover, computationally intensive waveform analysis can introduce latency, potentially limiting the achievable update rate of generated point clouds compared to simpler, single-return LiDAR systems.\cite{wallace2020full}.

\subsubsection{Feature-based Methods}
Feature-based denoising methods for LiDAR point clouds can be broadly classified into three categories: distance-based, intensity-based, and distance–intensity fusion methods \cite{park2025lidar}. 

Distance-based approaches rely solely on spatial relationships among points, making them effective for removing isolated outliers but often computationally intensive. These filters typically classify a point as noise if it lacks sufficient neighboring points—either within a fixed radius or among its $k$-nearest neighbors. This spatial isolation assumption holds for sparse phenomena such as drifting snowflakes~\cite{charron2018noising, zhou2024dcor}. 

However, under adverse weather conditions involving rain or fog, spurious returns often form dense curtains of micro-droplets or multiply scattered streaks. In such scenarios, each false return is likely to have many nearby neighbors, satisfying the spatial proximity criterion and thereby escaping removal. Since distance-based filters rely purely on geometric structure and ignore the low-intensity characteristics typical of weather-induced noise, they fail to suppress such points effectively. As a result, the LiDAR point cloud remains heavily contaminated, making these methods suboptimal for applications like level crossing obstacle detection under adverse weather conditions.

Intensity-based methods classify LiDAR points primarily according to reflected signal intensity, effectively differentiating environmental noise such as rain droplets or snowflakes from genuine object points due to the environmental noise's characteristically lower reflectance. These methods offer substantial advantages in computational speed and simplicity by employing straightforward intensity thresholding, which significantly accelerates processing compared to more computationally demanding spatial-only methods. 

Nevertheless, intensity-based methods inherently face limitations in scenarios where the intensity of object surfaces overlaps with noise points due to material properties or color reflectivity. Hence, accurate calibration and adaptive measures are essential to maintain robust performance across diverse environmental conditions.

Park \textit{et al.} proposed the Low-Intensity Outlier Removal (LIOR) filter, specifically tailored for snowy conditions. This filter distinguishes snow-induced noise points from object points based on intensity thresholds \cite{park2020fast}. The key strength of LIOR lies in its high computational efficiency. However, the approach occasionally faces issues in accurately distinguishing low-intensity object points from noise, necessitating a second spatial-density step to correct potential misclassifications.

In distance-intensity fusion methods for LiDAR denoising, spatial and intensity data are jointly analyzed to enhance noise removal accuracy beyond what individual methods can achieve. These fusion approaches effectively mitigate the weaknesses of single-feature methods, such as the inability of pure spatial filters to distinguish dense weather noise and the misclassification risk of pure intensity-based thresholds.

A notable example is the Reflectance and Geometrical Outlier Removal (RGOR) algorithm~\cite{han2023rgor}, which restores reflectance based on LiDAR optical properties and then refines results using local geometric context. This two-step process allows RGOR to distinguish snow or rain particles from legitimate object points. However, the method can misclassify low-reflectivity targets like glass as noise if geometry is not used, highlighting a trade-off between robustness and over-suppression.

Expanding on intensity-based frameworks, Han \textit{et al.} developed a combined intensity and spatial filtering method targeted at rainy weather. This method integrates a weighted edge-preserving filter for correcting distorted contours and employs low-intensity filtering to eliminate scattered noise from raindrops. Experimental results indicated notable improvements, with increased precision, a recall rate of 99.28\%, and enhanced target detection accuracy \cite{han2023denoising}. Similarly, Roriz \textit{et al.} introduced Dynamic Low-Intensity Outlier Removal (DIOR), which combines LIOR’s intensity criteria with DROR’s adaptive spatial search radius, significantly reducing false positives and boosting overall accuracy. Implemented on hardware platforms, DIOR achieves enhanced performance compared to other outlier removal solutions, while guaranteeing the real-time requirements, vital for automotive and level crossing applications \cite{DIOR2022}.

Other fusion techniques like the Dynamic Distance Intensity Outlier Removal (DDIOR) \cite{wang2022ddior} and the Low-Intensity Dynamic Statistical Outlier Removal (LIDSOR)~\cite{huang2023lidsor} extend this principle by adapting thresholds dynamically. DDIOR adjusts filters based on both distance and reflectance trends observed in snowy conditions, while LIDSOR employs gamma distribution modeling to statistically isolate weather-induced outliers. These methods show improved accuracy and environmental detail preservation over traditional filters, but still depend on proper parameter tuning and noise modeling assumptions to generalize well across different conditions.

\subsubsection{Learning-based Methods}
Learning-based methods are also widely applied in LiDAR imaging and object detection tasks. Although these approaches typically require large volumes of annotated data and significant computational resources, they often outperform traditional methods in terms of accuracy and robustness. However, a key limitation is that they are generally constrained to recognizing only objects that have been labeled during training.

One representative example is the TripleMixer network \cite{zhao2024triplemixer}, which introduces three specialized mixer layers geometry, frequency, and channel mixers, to denoise point clouds affected by rain, fog, and snow. This architecture preserves spatial structure while integrating frequency and contextual information, offering interpretability and strong generalization on datasets such as Weather-KITTI and Weather-NuScenes. Another approach, SAFDN~\cite{zhang2024improved}, extends the PP-LiteSeg segmentation backbone with WeatherBlock and Series Attention Fusion Modules to improve real-time segmentation and denoising. It achieves over 11\% performance gain compared to baseline models on the DENSE dataset, showing the benefit of combining multi-scale feature extraction with spatial-channel attention.

Several methods further enhance the trade-off between accuracy and efficiency. For instance, 3D-OutDet~\cite{raisuddin20243d} introduces a neighborhood-based convolution operation that significantly reduces memory usage (by 99.9\%) and execution time (by 82.8\%) with only minimal loss in segmentation accuracy compared to the baseline dense convolution model. Similarly, SMEDNet~\cite{seppanen2023multi} leverages multiple LiDAR echoes and a self-supervised learning framework to select signal-valid returns, effectively filtering out noise introduced by airborne particles without requiring point-level annotations.

In addition, Piroli et al.~\cite{piroli2023energy} propose an energy-based outlier detection method that reframes weather noise filtering as an unsupervised anomaly detection task. By assigning high energy to outliers and low energy to valid points, the method robustly detects and removes weather-induced noise. Here, the energy score is not related to LiDAR intensity but is derived from the network’s output logits, where lower energy indicates high confidence of being an inlier, and higher energy suggests anomalous, weather-induced points. Meanwhile, Heinzler et al.~\cite{heinzler2020cnn} present one of the earliest CNN-based denoising frameworks trained on controlled fog/rain datasets, offering improved robustness over traditional geometric filters.

Despite their strengths, learning-based methods still face challenges such as generalization to unseen weather phenomena and the scarcity of large, diverse annotated datasets. Balancing computational efficiency and robustness remains an ongoing area of research. In addition, the deployment of such data-driven approaches in safety-critical railway applications must comply with strict safety and certification standards, which poses further challenges related to explainability, verification, and regulatory approval.

\subsection{Sensor Fusion}
Sensor fusion is essential for robust obstacle detection in complex environments, as it facilitates the integration of complementary information from multiple sensing modalities. Based on the stage at which sensor data are combined, fusion strategies can be broadly classified into three categories: data-level fusion, feature-level fusion, and decision-level fusion. In this section, we provide a brief overview of representative methods within each category and discuss their respective advantages and limitations.

\subsubsection{Data-level Fusion}
Data-level fusion directly combines raw or minimally processed sensor data before higher-level feature extraction or decision-making, enabling systems to exploit the complementary nature of heterogeneous sensors. Common sensor pairings include LiDAR-Camera~\cite{caltagirone2018lidar}, Radar-Camera~\cite{chadwick2019distant}, and LiDAR-Radar~\cite{huang2024l4dr}. These combinations leverage spatial, temporal, and spectral diversity to improve detection reliability, especially under adverse conditions.

For instance, in LiDAR–camera fusion, raw point clouds are projected onto the camera plane to form dense 2D feature maps, allowing early-stage convolutional processing within a fully convolutional network (FCN). Early fusion strategies, such as concatenating LiDAR and RGB inputs prior to feature extraction~\cite{caltagirone2018lidar}—are computationally efficient and perform well under favorable conditions. However, these methods are vulnerable to asymmetric sensor degradation. Since early fusion combines sensor data at the input level, degradation in one modality can negatively impact the entire network’s performance.

Radar-camera fusion is particularly useful for distant object detection, where radar's long-range velocity measurements compensate for vision-based detection limits~\cite{chadwick2019distant}. In these methods, radar returns are projected onto the image space as additional channels (e.g., range and range-rate), enhancing small object detection at long distances. However, it should be noted that Doppler (range-rate) information is only effective for moving targets and does not directly contribute to detecting static obstacles, such as vehicles stalled at a level crossing. In such cases, fusion schemes need to rely more heavily on radar range, amplitude, or vision-based cues to ensure robust obstacle detection.

Recent work in LiDAR-4D Radar fusion~\cite{huang2024l4dr} addresses severe weather challenges by combining raw point clouds via modules such as Multi-Modal Encoders (MME) and Foreground-Aware Denoising (FAD). This form of early fusion reduces the influence of sensor-specific noise before higher-level representations are learned.

The main advantage of data-level fusion lies in its ability to exploit raw complementary data early, potentially leading to richer representations. However, it also poses challenges: sensors must be temporally and spatially calibrated, and misaligned or degraded inputs can propagate errors throughout the network. Moreover, differences in resolution, sampling density, and noise characteristics across modalities make it non-trivial to fuse data without introducing inconsistencies.

Despite these challenges, data-level fusion remains an essential approach for robust perception, especially when paired with noise-aware preprocessing and adaptive fusion modules to mitigate sensor degradation.

\subsubsection{Feature-Level Fusion}

Feature-level fusion aims to integrate complementary information from multiple sensor modalities—primarily camera, LiDAR, and radar—at the intermediate representation stage. Unlike data-level fusion that operates directly on raw signals, feature-level methods extract modality-specific features and subsequently merge them to form richer and more discriminative representations. A common approach involves aligning features spatially using calibration matrices and then fusing them via concatenation, attention mechanisms, or learned fusion layers.

For example, PoIFusion~\cite{deng2024poifusion} employs adaptive point-of-interest sampling to extract modality-specific features, followed by dynamic fusion blocks to combine multi-scale features from LiDAR and images. This mitigates misalignment and reduces the computational burden seen in global attention mechanisms. Similarly, AVOD~\cite{ku2018joint} aggregates features from bird's-eye-view LiDAR projections and front-view camera images using a shared region proposal network, achieving real-time performance while preserving small object localization. RRPN~\cite{nabati2019rrpn} leverages radar detections as anchors and augments camera-based proposals, demonstrating that radar-derived range and velocity can enhance spatial accuracy and reduce computation time. Radar-vision fusion is also explored in~\cite{chadwick2019distant}, where radar channels (range and Doppler) are projected into image space and combined with CNN-based features for improved small-object detection. Feature-level fusion's main strengths lie in its balance between robustness and computational efficiency, especially when sensor inputs are partially degraded. However, it remains sensitive to calibration inaccuracies and struggles with semantic inconsistency between sensor domains.

Overall, feature-level fusion offers a powerful framework to exploit the heterogeneity of multi-sensor systems, enabling enhanced object detection under challenging conditions, though its success is closely tied to effective feature alignment and adaptive fusion strategies.

\subsubsection{Decision-Level Fusion}

Decision-level fusion integrates the independent detection results from multiple sensors to arrive at a final decision through defined rules or probabilistic reasoning. This method is particularly advantageous in complex environments where each sensor alone may provide incomplete or uncertain data. In railway and autonomous driving contexts, decision-level fusion has been extensively explored using combinations of radar, camera, LiDAR, infrared, and fiber optic sensors.

In this fusion scheme, each sensor processes its data independently and provides object-level outputs such as target location, class, or velocity. These outputs are then combined using logic-based or probabilistic fusion methods. A prominent example is the use of radar-camera fusion where the Mahalanobis distance and Joint Probabilistic Data Association (JPDA) are employed to correlate detected targets across sensors based on positional similarity and dynamic state information~\cite{liu2022robust}. This approach enables robust target tracking and classification even in adverse weather, by exploiting radar’s robustness to poor visibility and camera’s superior classification ability.

Similarly, in railway perimeter intrusion detection, decision-level fusion systems combine outputs from radar, cameras, and infrared sensors. Rule-based voting strategies or weighted decision fusion are often applied to enhance reliability~\cite{shi2024survey}. While this high-level fusion is computationally efficient and resilient to individual sensor failures, it suffers from a major limitation: the reliance on earlier-stage detection quality. Information loss from discarding raw and intermediate data may reduce detection accuracy, especially when upstream sensors generate ambiguous results.

Despite this limitation, decision-level fusion is highly scalable and suitable for real-time applications. Its modularity also simplifies sensor redundancy handling and system maintenance, making it ideal for large-scale deployments in safety-critical applications such as high-speed rail and autonomous vehicles.

\subsection*{Summary}
Data-level fusion preserves the most raw information but requires precise calibration. Feature-level fusion balances accuracy and efficiency through intermediate representations. Decision-level fusion is computationally lightweight and modular, integrating individual sensor outputs at a high level. Each fusion strategy offers trade-offs in complexity and flexibility as summarized in table \ref{tab:fusion_summary}.

\begin{table*}[htbp]
\caption{Comparison of sensor fusion levels in obstacle detection applications. The table contrasts different fusion levels in terms of input format, computational complexity, synchronization requirements, and associated advantages and disadvantages.}
\centering
\scriptsize
\begin{tabular}{|l|l|c|c|p{6cm}|} 
\hline
\textbf{Level} & \textbf{Input Format} & \textbf{Comp. Complexity} & \textbf{Sync Req.} & \textbf{Pros / Cons} \\
\hline
Data-level Fusion & Raw/Preprocessed Data & High & High & 
+ Maximum detail, enables robust detection \newline 
-- Heavy computation, strict synchronization \\
\hline
Feature-level Fusion & Sensor Features & Medium & Medium & 
+ Balanced trade-off between accuracy and complexity \newline 
-- Some information loss compared to raw data \\
\hline
Decision-level Fusion & Detection/Classification Results & Low & Low & 
+ Simple integration, minimal synchronization \newline 
-- Limited contextual information, vulnerable to single-sensor errors \\
\hline
\end{tabular}
\label{tab:fusion_summary}
\end{table*}

\section{Conclusion and Future Outlook}

This review has provided a detailed survey of current sensor technologies and fusion strategies for obstacle detection at railway level crossings, with particular emphasis on performance under adverse weather conditions. Ensuring the safety and efficiency of railway transport requires robust, real-time, and weather-resilient detection systems capable of identifying a broad range of obstacles under diverse environmental conditions. However, each sensing modality analyzed in this paper—including inductive loops, cameras, radar, and LiDAR—exhibits inherent trade-offs that limit its performance when used in isolation.

Inductive loops, while cost-effective and resilient to adverse weather, are constrained by their reliance on metallic mass for detection. As such, they are ineffective in detecting non-metallic objects such as humans, animals, or lightweight debris. Their subsurface installation also imposes high operational costs during deployment or maintenance, often requiring disruptions to active railway service. Despite their weather resilience, inductive loops lack the capability to differentiate between harmless and hazardous objects under complex environmental noise. Currently, no effective method exists to extend their detection capability to non-metallic intrusions, limiting their applicability as a sole sensor in harsh outdoor environments.

Camera-based systems, including RGB, thermal, stereo, and event cameras, provide rich semantic information and are highly effective in detecting and classifying obstacles under clear weather conditions. Deep learning methods have enabled impressive advances in obstacle recognition, rail segmentation, and behavior inference. However, visual sensors remain highly vulnerable to environmental factors such as glare, rain, snow, fog, and night-time operation. Water droplets or condensation on the lens, image contrast degradation, and dynamic occlusions often result in significant drops in detection reliability. To mitigate these issues, traditional image enhancement techniques such as histogram equalization and contrast stretching have been complemented by advanced methods. For example, dehazing networks like DehazeNet and AOD-Net address fog-induced contrast loss, while rain and snow removal networks such as RESCAN and OMS-CSC use temporal and spatial redundancy to restore image integrity. Event-based cameras have also shown resilience in low-light and fast-motion conditions due to their high dynamic range and asynchronous readout. Despite these developments, many methods remain sensitive to failure cases and require significant computational resources and diverse training data to generalize across weather types.

Radar, particularly FMCW and MIMO configurations, offers inherent advantages in all-weather operation due to its use of longer wavelengths and active signal transmission. Its insensitivity to lighting and strong penetration ability make it a robust component in environments affected by rain, fog, and dust. Radar excels in detecting large and medium-sized moving objects and estimating their range and relative velocity. However, radar suffers from lower spatial resolution, significant interference from clutter, and limited capability to distinguish between object types. Several weather mitigation techniques have been developed, particularly for rain attenuation and multipath scattering. Analytical models such as the S-matrix formulation and stochastic propagation models help characterize radar distortions under adverse media. Covariance-based calibration methods leveraging Hermitian Toeplitz structures, such as the one proposed for DoA correction under rain, have shown improved resolution and estimation accuracy. Nonetheless, current radar mitigation strategies mostly remain model-driven and limited to specific geometric assumptions, with limited generalization to complex field settings or multiple-object scenarios.

LiDAR provides dense, accurate 3D point cloud data, offering unmatched spatial resolution for obstacle localization. In clear weather, LiDAR is highly effective at identifying and tracking both static and dynamic obstacles. Nevertheless, adverse weather conditions significantly affect point cloud integrity due to backscattering, multipath reflections, and intensity attenuation. To address this, several mitigation methods have been proposed. Feature-based filters such as LIOR, DROR, and DIOR classify noise using spatial density and intensity thresholds, and have shown effective removal of snowflakes or raindrops. Distance-intensity fusion methods like RGOR and DDIOR further improve performance by exploiting both spatial structure and reflectance information. Full-waveform LiDAR systems, by capturing continuous echo responses, are capable of distinguishing atmospheric backscatter from ground reflections, but require complex signal processing and incur high power and data overhead. Deep learning models like SAFDN and TripleMixer have demonstrated strong generalization across fog, rain, and snow conditions using spatial-frequency fusion and attention mechanisms. However, these models require extensive training and struggle with rare or rapidly changing weather scenarios. Thus, while LiDAR offers significant potential, its robustness depends heavily on tailored denoising algorithms and real-time processing capability.

\subsection*{Sensor Fusion: A Path Toward Robust Systems}

The growing recognition that no single sensor suffices under all conditions has led to increasing interest in multi-sensor fusion. Fusion strategies—classified as data-level, feature-level, and decision-level—enable systems to combine the complementary strengths of different sensors. 

Data-level fusion provides maximal information richness but demands high-precision spatial and temporal synchronization. It is also vulnerable to cascading errors from degraded sensors. Feature-level fusion, where modality-specific features are extracted before integration, offers a better trade-off between robustness and computational efficiency. Decision-level fusion, though information-sparse, provides superior fault tolerance, system modularity, and simplified maintenance—attributes critical for large-scale deployment.

In current literature, data-level fusion of LiDAR and camera has shown strong performance in clear conditions, while radar-camera fusion enhances performance at long ranges and during poor visibility. Feature-level fusion frameworks—such as PoIFusion and AVOD—have demonstrated resilience by selectively leveraging reliable features across modalities. Decision-level fusion, often used in safety-critical applications, utilizes probabilistic reasoning and rule-based logic to combine classification outputs from radar, vision, and infrared sensors.

However, challenges remain: effective fusion requires calibration, sensor failure diagnosis, domain adaptation, and real-time computation—tasks complicated by hardware diversity and environmental dynamics.

In addition to enhancing safety through reliable obstacle detection, several sensor technologies also enable secondary functionalities such as object classification and traffic characterization. For instance, vision-based systems and learning-augmented LiDAR frameworks can classify detected objects (e.g., vehicles, pedestrians, cyclists, or animals), allowing not only immediate risk assessment but also the collection of long-term statistics on traffic patterns at level crossings. Such information is valuable for railway operators and authorities in planning infrastructure upgrades, optimizing traffic management, and conducting safety evaluations. Thus, future sensor fusion frameworks should consider integrating these secondary features to simultaneously support both safety-critical detection and broader traffic monitoring objectives.

\subsection*{Closing Remarks}

In conclusion, obstacle detection at level crossings remains a complex yet urgent challenge. As railway networks modernize and traffic volumes increase, the need for intelligent, fail-safe detection systems becomes ever more critical. This survey has shown that while significant progress has been made in sensor hardware, signal processing, and learning algorithms, no single approach is sufficient on its own.

The most promising solution lies in sensor fusion: combining radar’s resilience, LiDAR’s precision, and vision’s semantic richness, complemented by novel sensing modalities and robust processing pipelines. Equally important are adaptive fusion algorithms, context-aware learning, comprehensive datasets, and scalable deployment architectures.

Ultimately, bridging the gap between laboratory research and field-ready systems will require interdisciplinary collaboration between signal processing, computer vision, transportation engineering, embedded systems, and regulatory policy. Such integrated efforts are essential to ensure the safety, efficiency, and reliability of future rail transport under all environmental conditions.

\bibliographystyle{IEEEtran}
\bibliography{ref}
\end{document}